\documentclass[prl,twocolumn,showpacs,floatfix]{revtex4}
\usepackage{graphics,epsfig,amsfonts,amssymb,amsmath}
\begin{document}

\title{Low magnetization and anisotropy in the antiferromagnetic state 
of undoped iron pnictides}
\author{E. Bascones}
\author{M.J. Calder\'on}
\author{B. Valenzuela}
\affiliation{Instituto de Ciencia de Materiales de Madrid,
ICMM-CSIC, Cantoblanco, E-28049 Madrid (Spain).}
\email{leni@icmm.csic.es,calderon@icmm.csic.es,belenv@icmm.csic.es}
\date{\today}
\begin{abstract}
We examine the magnetic phase diagram of iron pnictides using a five band model. For the intermediate values of the interaction expected to hold in the iron pnictides, we find a metallic low moment state characterized by antiparallel orbital magnetic moments. The anisotropy of the interorbital hopping amplitudes is the key to understanding this low moment state. This state accounts for the small magnetization measured in undoped iron pnictides and leads to the strong exchange anisotropy found in neutron experiments. Orbital ordering is concomitant with magnetism and produces the large $zx$ orbital weight seen at $\Gamma$ in photoemission experiments. 
\end{abstract}
\maketitle

One of the common features in most iron pnictides is the appearance of
unusual antiferromagnetism (AF) in the undoped compounds. In this state the
magnetic moments order with momentum ${\bf Q}=(\pi,0)$, namely 
antiferromagnetically in the $x$-direction and ferromagnetically in the 
$y$-direction\cite{cruz08}. The Curie temperatures 
are high $T_N \sim 130-200 K$, while 
the measured magnetic moment $m$
is small $m \sim 0.3-1.0 \mu_B$. 
A structural
transition at $T_s \geq T_N$ accompanies magnetism and the system 
shows metallic behavior in the magnetic state. 
In spite of the current hot
debate, the weak\cite{mazin08-2,raghu08} or strong
coupling\cite{si08,yildirim08} nature of magnetism is not clear
yet. Ab-initio calculations\cite{mazin08-2} generally report magnetic moments 
$m \geq 2 \mu_B$, much larger than experimentally measured.
Proposals to explain the small magnetic moment include 
frustration\cite{yildirim08}, weak
order\cite{japonesesmf09}, antiphase boundaries\cite{mazinnatphys09}, 
opposite orbital magnetizations\cite{cricchio09,rodriguez09} or the interplay 
between frustrated and non-frustrated bands\cite{valenti09}.
 
Unexpectedly, very anisotropic nearest neighbor exchange constants $J_y\ll
J_x$, with $J_y$ even slightly ferromagnetic, 
have been necessary to describe neutron scattering 
results\cite{zhaonatphys09}.  
Orbital ordering (OO) was proposed early on
within a Kugel-Khomskii description\cite{kruger09} and argued to
be behind both the strong anisotropy\cite{singh-2009,leeyinku09} and the
structural transition\cite{phillips09,leeyinku09}. 
Recent 
experiments\cite{shimojima10,sciencedavis10} 
have also been interpreted as manifestations of orbital ordering.
In particular ARPES\cite{shimojima10} experiments show that the Fermi 
surface at $\Gamma$ has a predominant $zx$ orbital component.

Here we study the magnetic phase diagram of a five band Hamiltonian at the
mean field level. Different magnetic regimes, mostly 
coexistent with OO, are found. At intermediate Hubbard
interaction, the system is AF and metallic and shows two different magnetic 
phases. In particular, we find a low moment (LM) phase which accounts 
for the small magnetization measured on iron pnictides. 
The LM arises as a consequence of partial cancellation 
of antiparallel orbital magnetic moments. 
The anisotropy of interorbital hoppings is the key to
explain this phase.  
To compare with neutron results, we estimate the anisotropy of the
exchange interactions finding it to be strong only in the low moment state. We also show that the
OO has a small contribution to the anisotropy of this state.  
On the other hand, the OO enhances  
the $zx$ orbital over the $yz$ orbital weight in the Fermi surface.

We start from the 5-band interacting Hamiltonian 
\begin{eqnarray}
\nonumber
& H &  = \sum_{i,j,\gamma,\beta,\sigma}T^{\gamma,\beta}_{i,j}c^\dagger_{i,\gamma,\sigma}c_{j,\beta,\sigma}+h.c. 
+ U\sum_{j,\gamma}n_{j,\gamma,\uparrow}n_{j,\gamma,\downarrow}
\\ \nonumber & +&  U'\sum_{j,\gamma>\beta,\sigma,\tilde{\sigma}}n_{j,\gamma,\sigma}n_{j,\beta,\tilde{\sigma}}
+\frac{J}{2}\sum_{j,\gamma \neq
  \beta,\sigma,\tilde{\sigma}}c^\dagger_{j,\gamma,\sigma}c^\dagger_{j,\beta,\tilde{\sigma}}c_{j,\gamma,\tilde{\sigma}}c_{j,\beta,\sigma}
\\
& + &  J'\sum_{j,\gamma\neq
  \beta}c^\dagger_{j,\gamma,\uparrow}c^\dagger_{j,\gamma,\downarrow}c_{j,\beta,\downarrow}c_{j,\beta,\uparrow}  \,.
\label{eq:hamiltoniano}
\end{eqnarray}
Here $i,j$ label the Fe sites in the Fe unit cell. $\gamma$ and $\beta$ refer to the
five Fe-d orbitals $yz$, $zx$, $xy$, $3z^2-r^2$ and $x^2-y^2$ included in the
model, and $\sigma$ to the spin. 
$x$ and $y$ axis are directed along the Fe bonds.
The kinetic energy term includes hopping up to second neighbors 
with the hopping amplitudes calculated within the Slater-Koster
framework\cite{slater54} as
detailed in Ref.\cite{nosotrasprb09}. Both direct Fe-Fe and indirect (via As)
hoppings determine the magnitude of the hopping amplitudes. This tight binding model\cite{nosotrasprb09} 
gives good account of  the band structure found in density 
functional theory, including the orbital 
content of the bands, with a reduced number of fitting parameters.  
The hopping amplitudes depend on the 
angle $\alpha$ formed by the Fe-As bonds and the Fe-plane.
\begin{figure}
\leavevmode
\includegraphics[clip,width=0.4\textwidth]{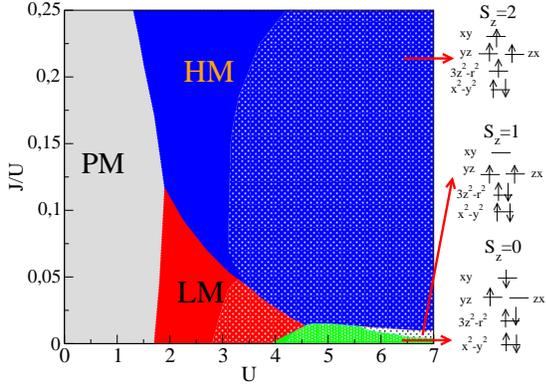}
\caption
{(Color online) Calculated $(\pi,0)$-mean field magnetic phase diagram of the five 
band Hamiltonian in Eq.~\ref{eq:hamiltoniano}.
Patterned areas (for $U \gtrsim 3$) correspond to gapped states. The following color code applies:
gray corresponds to the paramagnetic (PM) state;
red to a low magnetic moment (LM) state showing orbital
magnetizations $m_{\gamma}$ with opposite signs; blue to high magnetization (HM) with parallel $m_{\gamma}$ which corresponds at large $U$ to an $S_z=2$ state; white is an intermediate
$S_z=1$ moment state; and green is the $S_z=0$ state. The $S_z=2$, $S_z=1$ and $S_z=0$ states are illustrated from top to bottom on the right. Magnetic moment $m$ and gap are assumed to be finite when
larger than $0.001$.}
\label{fig:phasediag}
\end{figure}
In the following, $\alpha=35.3^o$ corresponding to the regular tetrahedra 
is used, except otherwise indicated. The interacting part of the Hamiltonian  
includes the intraorbital $U$ and the interorbital $U'$ interactions, 
as well as the
Hund's coupling $J$ and pair hopping $J'$ terms. The pair hopping interaction
is written for completeness, but it does not enter in the mean-field
approximation used below. Energies are given in units of
$(pd\sigma)^2/|\epsilon_d-\epsilon_p|\sim 1$ eV, with $pd\sigma$ the $\sigma$
overlap between the Fe-d and As-p orbitals and 
$|\epsilon_d-\epsilon_p|$ their energy difference\cite{nosotrasprb09}. 
We assume that AF takes place with $\bf{Q}=(\pi,0)$ momentum, as
experimentally observed, and treat the Hamiltonian at the mean field level
keeping only the spin and orbital-diagonal average 
terms\cite{daghofer09}
\begin{equation}
n_{\gamma}=\sum_{k,\sigma} \langle c^\dagger_{k,\gamma,\sigma}
c_{k,\gamma,\sigma}\rangle,\, m_{\gamma}=\sum_{k,\sigma} \sigma \langle c^\dagger_{k+\bf{Q},\gamma,\sigma}
c_{k,\gamma,\sigma}\rangle \, ,
\label{eq:hamil}
\end{equation}
where $k$ runs over the Fe Brillouin zone and $\sigma=\pm 1$.
This corresponds
to a magnetic moment $m=\sum_\gamma m_\gamma$, in units of $\mu_B$.
We assume that the relation $U'=U-2J$ from
rotational invariance\cite{castellani78} holds 
and study the phase diagram in the
$J/U$ versus $U$ space. Ferro-orbital ordering, 
denoted simply by OO in the following, and AF with momentum {\bf Q} are the only symmetry 
breaking states allowed in our calculation. 
We focus on undoped systems with $n=\sum_\gamma n_\gamma=6$.

\begin{figure}
\leavevmode
\includegraphics[clip,width=0.45\textwidth]{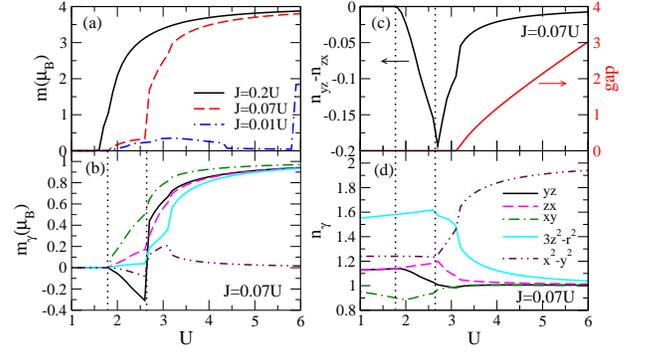}
\caption{
(Color online) (a) Total magnetic moment for three different values of $J$ as a function of $U$. (b) Orbital magnetic moments $m_{\gamma}$, (c) OO $n_{yz}-n_{zx}$ and gap, and (d) orbital filling $n_{\gamma}$ as a function of $U$ for $J/U=0.07$. The low magnetic moment state arising for $2 \leq U \leq 2.8$ is due to the partial cancellation of opposite magnetic moments on different orbitals.}
\label{fig:momentop07}
\end{figure}

Fig.~\ref{fig:phasediag}  displays the phase diagram with the paramagnetic (PM) state shown in 
gray. 
Different magnetic phases with momentum ${\bf Q}$
are represented in green, red, blue and white and correspond
to states with vanishing ($S_z =0$), low (LM), high (HM) and intermediate, 
denoted
$S_z=1$, magnetic moment, respectively. 
Shaded areas (for $U\gtrsim 3$) correspond to insulating states with a gap at the Fermi level. 
The PM state, which shows no OO, survives up to a
J-dependent critical value of $U$, $U_c(J)\sim 1.8$.   
The high value of
$U_c(J)$ is not expected in a nesting scenario and 
suggests a strong coupling origin of AF. 
Moreover, the magnetic states found for intermediate values
of $U\sim 2-3$ are metallic. 

\begin{figure}
\leavevmode
\includegraphics[clip,width=0.45\textwidth]{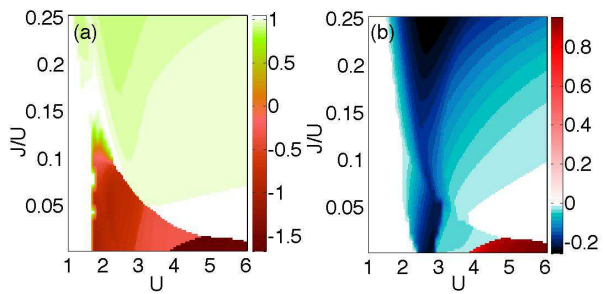}
\caption{
(Color online) (a) Anisotropy $J^y_{\rm eff}/J^x_{\rm eff}$ of the magnetic exchange from a strong coupling
approach and (b) orbital ordering $n_{yz}-n_{zx}$ as a function of $U$ and
$J$. The anisotropy appears mostly within the $S_z=0$ and low moment states 
(green and red (LM) regions in Fig.\ref{fig:phasediag}). The orbital ordering 
accompanies the magnetization within a wide range of parameters 
but is not correlated with the exchange anisotropy. }
\label{fig:anisotropy}
\end{figure}

For large $U$ the system evolves towards the atomic limit (see the sketches in
Fig.~\ref{fig:phasediag}) with a well defined filling of orbital 
and spin states. 
For $J \sim 0.01U$, an intermediate
magnetic moment $S_z=1$  and insulating state appears. In this state there is no
OO ($n_{3z^2-r^2}, n_{x^2-y^2} \sim 2,n_{yz},n_{zx}
\sim 1$) and  $m_{yz},  m_{zx} \sim 1 \mu_B$.
When $J$ increases, a spin from one of the doubly occupied states is promoted
to the $xy$ orbital and we recover the $S_z=2$ atomic limit with $ n_{x^2-y^2} \sim
2$,  $n_{z^2-r^2},n_{yz},n_{zx},n_{xy}\sim 1$ and very 
weak OO [see Fig.~\ref{fig:anisotropy}(b)]. In this 
state all the spins in
  the half occupied states are parallel. 
This tendency at large $U$ can be
appreciated 
in  Fig.~\ref{fig:momentop07}. The HM state survives 
when reducing $U$ for sufficiently large
values of $J/U$.
At intermediate values of $U$ the system is metallic and shows OO. 
 From Fig.~\ref{fig:momentop07} (a) 
it is apparent that, within this HM state, 
achieving the small magnetization $m< 1 \mu_B$ reported 
experimentally is only possible by fine-tuning, 
as in the weak order state discussed in Ref.~\cite{japonesesmf09}.

The $S_z=1$ and $S_z=2$ atomic states have been widely used as starting points 
in previous works, where the Hund's rule is assumed to hold. As shown in 
Fig.~\ref{fig:phasediag} for large $U$ and $J\simeq 0$, an $S_z=0$ state with a
strong and positive OO $n_{yz} >> n_{zx}$ appears. In this
state Hund's rule is violated. The electrons in the $xy$ and $yz$ orbitals have
antiparallel spins, see sketch at the right.
Most importantly, at intermediate $U$ and not too large $J/U$, we find an LM phase 
(with $m<1 \mu_B$) with negative OO $n_{yz}< n_{zx}$ 
in which the individual orbital 
magnetizations  have opposite signs.
In this state $m_{yz}, m_{x^2-y^2} <0$, while $m_{zx}, m_{xy}
>0$ ($m_{3z^2-r2} \sim 0$).  This is shown in Fig.\ref{fig:momentop07} (b) for
$J/U=0.07$ within the range $1.8<U<2.7$ (between the vertical dotted
lines). The low or almost vanishing total magnetic moment in these two 
states arises 
due to the partial cancellation of otherwise relatively strong magnetic 
moments in close to half-filling orbitals.

An AF solution with opposite orbital magnetizations, stabilized by the
formation of large multipoles of the spin magnetization, 
was previously found in an ab-initio calculation\cite{cricchio09} for 
LaOFeAs. Violation of Hund's rule in the pnictides has also been 
discussed within the context of a two-orbital Heisenberg model\cite{rodriguez09} as a 
consequence of large interorbital exchange
$J_{\gamma,\beta}$. In the latter work, isotropic $J_{\gamma,\beta}$ were used and the columnar AF order was not found. 
As discussed below, in our model the stability of the 
$S_z=0$ and LM states can be understood
 within a strong coupling
point of view by considering the 
{\it anisotropy of the interorbital exchange 
interactions}  $J_{\gamma,\beta}\propto t^2_{\gamma,\beta}$ which override
Hund's rule. 
\begin{figure}
\leavevmode
\includegraphics[clip,width=0.45\textwidth]{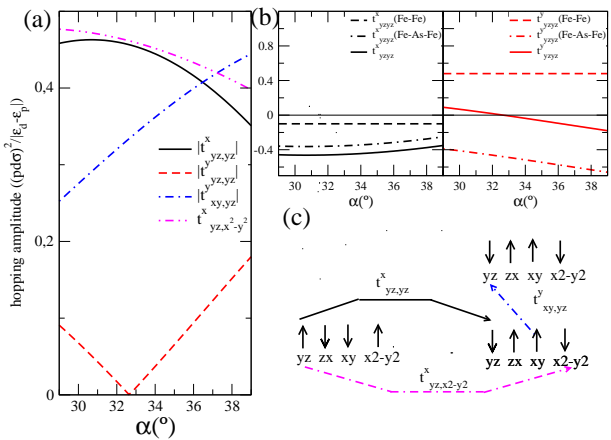}
\caption{
(Color online) (a) Dependence on $\alpha$ of the first nearest neighbors hopping amplitudes
mostly responsible for the stability of the low-magnetic moment state and the
orbital ordering. For other hopping amplitudes see~\cite{nosotrasprb09}. 
(b) Direct Fe-Fe and indirect (via As) contributions 
to $t^x_{yz,yz}$ and $t^y_{yz,yz}$ showing a strong anisotropy. (c)
Sketch of the low magnetic moment state on three neighboring sites of the Fe plane and the largest hoppings involved in its stabilization.}
\label{fig:hoppings}
\end{figure}

In an AF 
state with momentum ${\bf Q}=(\pi,0)$, each $m_{\gamma}$ changes sign
between nearest neighbors along the $x$ direction, but not along the $y$
direction. As shown in Fig.~\ref{fig:hoppings}(a), the hopping 
between $yz$ and $x^2-y^2$ along
the $x$ direction $|t^x_{yz,x^2-y^2}|$ is large, while it vanishes along the 
$y$ direction. Therefore, parallel magnetic order of $yz$ and $x^2-y^2$ is 
favored as the system gains AF energy along $x$ without any
cost in energy associated with the ferromagnetic ordering along $y$. On the 
contrary, the hopping between $yz$ and $xy$ is large along the $y$ direction 
$|t^y_{yz,xy}|$ and vanishes along
$x$. This implies that parallel ordering of the $yz$ and $xy$ magnetic moments would cost exchange
energy along $y$, while there would be no gain whatsoever along $x$. 
A configuration with opposite signs of $m_{yz}$ and
$m_{xy}$ shows relative AF ordering of these two magnetic
moments along $y$ (see sketch in Fig.~\ref{fig:hoppings}) saving exchange interaction. From symmetry, $|t^y_{zx,x^2-y^2}|= |t^x_{yz,x^2-y^2}|$ and
$|t^x_{zx,xy}|=|t^y_{yz,xy}|$ which favors antiparallel orientation between $m_{zx}$ and $m_{x^2-y^2}$ and parallel orientation between $m_{zx}$ and $m_{xy}$. 
Interactions involving $3z^2-r^2$ are more frustrated, $n_{3z^2-r^2}\sim 2$ 
and therefore its magnetization is
small in the LM state. At large $J/U$, Hund's rule dominates and the 
lowest energy correspond to the HM state.

Negative OO appears both in the LM and HM
regimes [see Fig.~\ref{fig:anisotropy}(b)].
This produces a splitting of $zx$ and $yz$ bands at $\Gamma$ leading to 
larger weight of $zx$ at the Fermi surface (see supplementary material\cite{supp}) in agreement with ARPES
experiments\cite{shimojima10}.

As discussed previously by Lee, Yin and Ku~\cite{leeyinku09} based on 
ab-initio calculations, OO originates in the 
anisotropy of the $yz$ ($zx$) intraorbital first nearest neighbor 
hopping which favors a large (small) magnetic moment in the 
$yz$ ($zx$) orbital. Unexpectedly, 
$t^x_{yz,yz}>t^y_{yz,yz}$ ($t^y_{zx,zx}>t^x_{zx,zx}$). 
This relationship
is opposite to the one used in early proposals of OO in iron 
pnictides\cite{kruger09,singh-2009}. 
Our tight-binding model\cite{nosotrasprb09} does not only reproduce Lee, Yin and
Ku~\cite{leeyinku09} results, but it also allows to understand its origin. From Fig.~\ref{fig:hoppings}(b)
it can be seen that the small value of 
$t^y_{yz,yz}$ ($=t^x_{zx,zx}$) comes from the  cancellation of large 
direct Fe-Fe and indirect (via As) contributions 
with opposite sign, which add with the same sign for 
$t^x_{yz,yz}$ ($=t^y_{zx,zx}$).

In order to compare with neutron experiments\cite{zhaonatphys09} 
we make connection 
with a Heisenberg model and estimate the anisotropy 
of the exchange interactions. 
Within a strong coupling approach, 
$J^{x,y}_{\gamma,\beta}\propto (t^{x,y}_{\gamma,\beta})^2$, and assuming 
$J_{\rm eff}^{x,y}=\sum_{\gamma,\beta} J^{x,y}_{\gamma,\beta} \vec
{s}_{\gamma,i} \vec{s}_{\beta,j} /(\vec{S}_i \vec{S}_j)$ with
  $s_{\gamma,j}$ and $S_j$ the orbital and total spin at site $i$, we have
calculated
the ratio $J^y_{\rm eff}/J^x_{\rm eff}$. 
The results for the anisotropy are shown 
as a function of $U$ and $J/U$ in Fig.~\ref{fig:anisotropy} (a). 
Large anisotropy $J^y_{\rm eff} \ll J^x_{\rm eff}$, including negative values of 
 $J^y_{\rm eff}/J^x_{\rm eff}$, is only found in the LM state.  A value comparable to the experimental\cite{zhaonatphys09} $J^y_{\rm eff}/J^x_{\rm eff} \sim -0.11$ is found in the LM state close to the transition to the HM state.  In the HM state, the largest calculated anisotropy corresponds to $J^y_{\rm eff}/J^x_{\rm eff} \sim 0.88$, very far from the experimental value.
Comparing with Fig.~\ref{fig:anisotropy} (b), it is apparent that the 
exchange anisotropy in Fig.~\ref{fig:anisotropy} (a) is not strongly related to OO 
(which appears in most of the phase diagram), 
but to the presence of antiparallel orbital magnetizations, 
except for the insulating strong coupling solution $S_z=0$ where both
effects are present.
This is easily understood as the contribution of products 
$\vec s_\gamma \vec s_\beta$ with positive and negative 
sign  cancel when summed up. 
Following a similar procedure we have also estimated the ratio 
of the exchange between second and first neighbors
$J^{(2)}_{\rm eff}/J^x_{\rm eff}$.
Its value, not shown, 
is around $1/2$ in HM state but decreases in the LM phase.  This is also consistent with the experimental\cite{zhaonatphys09} $J^{(2)}_{\rm eff}/J^x_{\rm eff} \sim 0.38$. This 
suggests that second nearest
neighbors are important to stabilize the HM, but not the LM state.

Several works have emphasized the influence of the Fe-As angle 
$\alpha$ on determining the properties of iron 
pnictides\cite{iyo08}. 
We have calculated the phase diagram for 
$\alpha=37.2^o$ (elongated tetrahedra) and for $\alpha=29.9^o$ (squashed
tetrahedra as found in LaFePO). For the elongated case
the phase diagram is very similar to the one discussed
here for the regular tetrahedra.
On the other hand, for $\alpha=29.9^o$ 
the low magnetic moment is less stable and
reduced to a very small portion of the phase diagram. 
It is interesting to note that AF is absent in LaFePO.
The worse stability of the LM state for squashed tetrahedra 
can be understood
by looking at the angle dependence of the hopping parameters (see
Fig.~\ref{fig:hoppings}(a)): $|t^y_{yz,xy}|$ which helps stabilize the 
LM phase 
strongly decreases with decreasing $\alpha$. 

In conclusion, we have studied the mean field magnetic phase 
diagram of a 
five orbital model for the iron pnictides. Several magnetic phases appear for
  different values of Hubbard and Hund interactions. OO 
is present for a wide range of parameters  but does not seem to
have a strong effect on the anisotropy of the exchange interactions
in the metallic region. 
A  metallic low magnetization state with antiparallel 
orbital magnetic moments is found 
for intermediate values of $U$. 
It is stabilized by the 
anisotropy of the interorbital hoppings, ultimately related to the 
symmetry of the orbitals and the tetrahedral coordination 
of the As atoms. 
This state is consistent with the measured small magnetic 
moment\cite{cruz08}, 
the large weight of $zx$ found in the Fermi Surface 
around $\Gamma$\cite{shimojima10},
and the anisotropy of the exchange interactions\cite{zhaonatphys09}.

Our results uncover antiparallel orbital magnetizations as a new source 
of anisotropy connecting not previously
related experiments: the anisotropy measured with neutron scattering and the
observed low magnetic moment.  
They suggest a strongly correlated origin of the magnetic state 
different from both the nesting scenario and the usual 
Heisenberg description in terms of the atomic moment
and point to the need of describing correctly the individual exchange 
interactions between the orbital magnetic moments.
They also stress the importance of including all five Fe 
d-orbitals in the description of the 
pnictides and question the
validity of band models which mimic the band structure and 
Fermi surface with
less orbitals as the hopping amplitudes are expected to differ
considerably from the correct ones.    

To the best of our knowledge, we are not aware of any technique that can make a straightforward direct measurement of the low magnetic moment state that we find. However, we expect it to show up indirectly as we have already seen in the anisotropy of the exchange interaction observed with neutron experiments.

We have benefited from conversations with M. Daghofer, F. Yndurain, F. Mompe\'an, V. Scagnoli, N. Nemes, C. Prieto, N. Men\'endez, and M.A. Laguna. We
  acknowledge funding from Ministerio de Ciencia e Innovaci\'on through Grants
  No. FIS 2008-00124/FIS, and No. FIS2009-08744 and Ram\'on y Cajal contract,
  and from Consejer\'ia de Educaci\'on de la Comunidad Aut\'onoma de Madrid and
  CSIC through Grants No. CCG08-CSIC/ESP-3518, PIE-200960I033
  and PIE-200960I180.


\begin{thebibliography}{29}
\expandafter\ifx\csname natexlab\endcsname\relax\def\natexlab#1{#1}\fi
\expandafter\ifx\csname bibnamefont\endcsname\relax
  \def\bibnamefont#1{#1}\fi
\expandafter\ifx\csname bibfnamefont\endcsname\relax
  \def\bibfnamefont#1{#1}\fi
\expandafter\ifx\csname citenamefont\endcsname\relax
  \def\citenamefont#1{#1}\fi
\expandafter\ifx\csname url\endcsname\relax
  \def\url#1{\texttt{#1}}\fi
\expandafter\ifx\csname urlprefix\endcsname\relax\def\urlprefix{URL }\fi
\providecommand{\bibinfo}[2]{#2}
\providecommand{\eprint}[2][]{\url{#2}}

\bibitem[{\citenamefont{de~la Cruz et~al.}(2008)\citenamefont{de~la Cruz,
  Huang, Lynn, Li, Ratcliff, Zarestky, Mook, Chen, Luo, Wang et~al.}}]{cruz08}
\bibinfo{author}{\bibfnamefont{C.}~\bibnamefont{de~la Cruz}} 
\bibnamefont{et~al.},
  \bibinfo{journal}{Nature} \textbf{\bibinfo{volume}{453}},
  \bibinfo{pages}{899} (\bibinfo{year}{2008}).

\bibitem[{\citenamefont{Mazin et~al.}(2008)\citenamefont{Mazin, Johannes,
  Boeri, and Koepernik}}]{mazin08-2}
\bibinfo{author}{\bibfnamefont{I.}~\bibnamefont{Mazin}} 
\bibnamefont{et~al.},
  \bibinfo{journal}{Phys. Rev. B} \textbf{\bibinfo{volume}{78}},
  \bibinfo{pages}{085104} (\bibinfo{year}{2008}).

\bibitem[{\citenamefont{Raghu et~al.}(2008)\citenamefont{Raghu, Qi, Liu,
  Scalapino, and Zhang}}]{raghu08}
\bibinfo{author}{\bibfnamefont{S.}~\bibnamefont{Raghu}} 
\bibnamefont{et~al.},
  \bibinfo{journal}{Phys. Rev. B} \textbf{\bibinfo{volume}{77}},
  \bibinfo{pages}{220503} (\bibinfo{year}{2008}).

\bibitem[{\citenamefont{Si and Abrahams}(2008)}]{si08}
\bibinfo{author}{\bibfnamefont{Q.}~\bibnamefont{Si}} \bibnamefont{and}
  \bibinfo{author}{\bibfnamefont{E.}~\bibnamefont{Abrahams}},
  \bibinfo{journal}{Phys. Rev. Lett.} \textbf{\bibinfo{volume}{101}},
  \bibinfo{pages}{076401} (\bibinfo{year}{2008}).

\bibitem[{\citenamefont{Yildirim}(2008)}]{yildirim08}
\bibinfo{author}{\bibfnamefont{T.}~\bibnamefont{Yildirim}},
  \bibinfo{journal}{Physical Review Letters} \textbf{\bibinfo{volume}{101}},
  \bibinfo{pages}{057010} (\bibinfo{year}{2008}).

\bibitem[{\citenamefont{Kaneshita et~al.}(2009)\citenamefont{Kaneshita,
  Morinari, and Tohyama}}]{japonesesmf09}
\bibinfo{author}{\bibfnamefont{E.}~\bibnamefont{Kaneshita}} 
\bibnamefont{et~al.},
  \bibinfo{journal}{Phys. Rev. Lett.} \textbf{\bibinfo{volume}{103}},
  \bibinfo{pages}{247202} (\bibinfo{year}{2009}).

\bibitem[{\citenamefont{Mazin and Johannes}(2009)}]{mazinnatphys09}
\bibinfo{author}{\bibfnamefont{I.}~\bibnamefont{Mazin}} \bibnamefont{and}
  \bibinfo{author}{\bibfnamefont{M.}~\bibnamefont{Johannes}},
  \bibinfo{journal}{Nature Physics} \textbf{\bibinfo{volume}{5}},
  \bibinfo{pages}{141} (\bibinfo{year}{2009}).

\bibitem[{\citenamefont{Cricchio et~al.}(2009)\citenamefont{Cricchio, Granas,
  and Nordstrom}}]{cricchio09}
\bibinfo{author}{\bibfnamefont{F.}~\bibnamefont{Cricchio}},
  \bibinfo{author}{\bibfnamefont{O.}~\bibnamefont{Granas}}, \bibnamefont{and}
  \bibinfo{author}{\bibfnamefont{L.}~\bibnamefont{Nordstrom}},
  \bibinfo{journal}{arXiv:0911.1342}  (\bibinfo{year}{2009}).

\bibitem[{\citenamefont{Rodriguez and Rezayi}(2009)}]{rodriguez09}
\bibinfo{author}{\bibfnamefont{J.~P.} \bibnamefont{Rodriguez}}
  \bibnamefont{and} \bibinfo{author}{\bibfnamefont{E.~H.}
  \bibnamefont{Rezayi}}, \bibinfo{journal}{Phys. Rev. Lett.}
  \textbf{\bibinfo{volume}{103}}, \bibinfo{pages}{097204}
  (\bibinfo{year}{2009}).

\bibitem[{\citenamefont{Lee et~al.}(2009{\natexlab{a}})\citenamefont{Lee,
  Zhang, Jeschke, and Valenti}}]{valenti09}
\bibinfo{author}{\bibfnamefont{H.}~\bibnamefont{Lee}} 
\bibnamefont{et~al.},
  \bibinfo{journal}{arXiv:0912.4024}  (\bibinfo{year}{2009}{\natexlab{a}}).

\bibitem[{\citenamefont{Zhao et~al.}(2009)\citenamefont{Zhao, Adroja, Yao,
  Bewley, Li, Wang, Wu, Chen, Hu, and Dai}}]{zhaonatphys09}
\bibinfo{author}{\bibfnamefont{J.}~\bibnamefont{Zhao}} 
\bibnamefont{et~al.},
  \bibinfo{journal}{Nature Physics} \textbf{\bibinfo{volume}{5}},
  \bibinfo{pages}{555} (\bibinfo{year}{2009}).

\bibitem[{\citenamefont{Kr\"uger et~al.}(2009)\citenamefont{Kr\"uger, Kumar,
  Zaanen, and van~den Brink}}]{kruger09}
\bibinfo{author}{\bibfnamefont{F.}~\bibnamefont{Kr\"uger}} 
\bibnamefont{et~al.},
  \bibinfo{journal}{Phys. Rev. B} \textbf{\bibinfo{volume}{79}},
  \bibinfo{pages}{054504} (\bibinfo{year}{2009}).

\bibitem[{\citenamefont{Singh}(2009)}]{singh-2009}
\bibinfo{author}{\bibfnamefont{R.~R.~P.} \bibnamefont{Singh}},
  \bibinfo{journal}{arXiv:0903.4408}  (\bibinfo{year}{2009}).

\bibitem[{\citenamefont{Lee et~al.}(2009{\natexlab{b}})\citenamefont{Lee, Yin,
  and Ku}}]{leeyinku09}
\bibinfo{author}{\bibfnamefont{C.-C.} \bibnamefont{Lee}},
  \bibinfo{author}{\bibfnamefont{W.-G.} \bibnamefont{Yin}}, \bibnamefont{and}
  \bibinfo{author}{\bibfnamefont{W.}~\bibnamefont{Ku}}, \bibinfo{journal}{Phys.
  Rev. Lett.} \textbf{\bibinfo{volume}{103}}, \bibinfo{pages}{267001}
  (\bibinfo{year}{2009}{\natexlab{b}}).

\bibitem[{\citenamefont{Lv et~al.}(2009)\citenamefont{Lv, Wu, and
  Phillips}}]{phillips09}
\bibinfo{author}{\bibfnamefont{W.}~\bibnamefont{Lv}},
  \bibinfo{author}{\bibfnamefont{J.}~\bibnamefont{Wu}}, \bibnamefont{and}
  \bibinfo{author}{\bibfnamefont{P.}~\bibnamefont{Phillips}},
  \bibinfo{journal}{Phys. Rev. B} \textbf{\bibinfo{volume}{80}},
  \bibinfo{pages}{224506} (\bibinfo{year}{2009}).

\bibitem[{\citenamefont{Shimojima et~al.}(2010)\citenamefont{Shimojima,
  Ishizaka, Ishida, Katayama, Ohgushi, Kiss, Okawa, Togashi, Wang, Chen
  et~al.}}]{shimojima10}
\bibinfo{author}{\bibfnamefont{T.}~\bibnamefont{Shimojima}} 
\bibnamefont{et~al.},
 \bibinfo{journal}{Phys. Rev. Lett.}
  \textbf{\bibinfo{volume}{104}}, \bibinfo{pages}{057002}
  (\bibinfo{year}{2010}).

\bibitem[{\citenamefont{Chuang et~al.}(2010)\citenamefont{Chuang, Allan, Lee,
  Xie, Ni, Bud'ko, Boebinger, Canfield, and Davis}}]{sciencedavis10}
\bibinfo{author}{\bibfnamefont{T.-M.} \bibnamefont{Chuang}}
  \bibinfo{author}{et al.},
  \bibinfo{journal}{Science} \textbf{\bibinfo{volume}{327}},
  \bibinfo{pages}{181} (\bibinfo{year}{2010});
\bibinfo{author}{\bibfnamefont{S.~H.} \bibnamefont{Lee}} 
  \bibnamefont{et al.}, \bibinfo{journal}{arXiv:0912.3205}
  (\bibinfo{year}{2009}{\natexlab{c}});
\bibinfo{author}{\bibfnamefont{S.}~\bibnamefont{Nandi}}
  \bibnamefont{et al.}, \bibinfo{journal}{Phys. Rev. Lett.} \textbf{\bibinfo{volume}{104}},
  \bibinfo{pages}{057006} (\bibinfo{year}{2010});
\bibinfo{author}{\bibfnamefont{R.~M.} \bibnamefont{Fernandes}} \bibnamefont{et al.},
  \bibinfo{journal}{arXiv:0911.3084}  (\bibinfo{year}{2009}); 
  \bibinfo{author}{\bibfnamefont{J.~G.} \bibnamefont{Analytis}}  \bibnamefont{et al.},  
  \bibinfo{journal}{arXiv:0911.3878}  (\bibinfo{year}{2009}).






\bibitem[{\citenamefont{Slater and Koster}(1954)}]{slater54}
\bibinfo{author}{\bibfnamefont{J.}~\bibnamefont{Slater}} \bibnamefont{and}
  \bibinfo{author}{\bibfnamefont{G.}~\bibnamefont{Koster}},
  \bibinfo{journal}{Phys. Rev.} \textbf{\bibinfo{volume}{94}},
  \bibinfo{pages}{1498} (\bibinfo{year}{1954}).

\bibitem[{\citenamefont{Calder\'on et~al.}(2009)\citenamefont{Calder\'on,
  Valenzuela, and Bascones}}]{nosotrasprb09}
\bibinfo{author}{\bibfnamefont{M.~J.} \bibnamefont{Calder\'on}},
  \bibinfo{author}{\bibfnamefont{B.}~\bibnamefont{Valenzuela}},
  \bibnamefont{and} \bibinfo{author}{\bibfnamefont{E.}~\bibnamefont{Bascones}},
  \bibinfo{journal}{Phys. Rev. B} \textbf{\bibinfo{volume}{80}},
  \bibinfo{pages}{094531} (\bibinfo{year}{2009}).

\bibitem[{\citenamefont{Yu et~al.}(2009)\citenamefont{Yu, Trinh, Moreo,
  Daghofer, Riera, Haas, and Dagotto}}]{daghofer09}
\bibinfo{author}{\bibfnamefont{R.}~\bibnamefont{Yu}} 
\bibnamefont{et~al.},
  \bibinfo{journal}{Phys. Rev. B} \textbf{\bibinfo{volume}{79}},
  \bibinfo{pages}{104510} (\bibinfo{year}{2009});
\bibinfo{author}{\bibfnamefont{M.}~\bibnamefont{Daghofer}} 
\bibnamefont{et~al.},
  \bibinfo{journal}{Phys. Rev. B} \textbf{\bibinfo{volume}{81}},
  \bibinfo{pages}{014511} (\bibinfo{year}{2010});
\bibinfo{author}{\bibfnamefont{S.}~\bibnamefont{Zhou}} \bibnamefont{and}
  \bibinfo{author}{\bibfnamefont{Z.}~\bibnamefont{Wang}},
  \bibinfo{journal}{arXiv:0910.2707}  (\bibinfo{year}{2009}).

\bibitem[{\citenamefont{Castellani et~al.}(1978)\citenamefont{Castellani,
  Natoli, and Ranninger}}]{castellani78}
\bibinfo{author}{\bibfnamefont{C.}~\bibnamefont{Castellani}},
  \bibinfo{author}{\bibfnamefont{C.~R.} \bibnamefont{Natoli}},
  \bibnamefont{and}
  \bibinfo{author}{\bibfnamefont{J.}~\bibnamefont{Ranninger}},
  \bibinfo{journal}{Phys. Rev. B} \textbf{\bibinfo{volume}{18}},
  \bibinfo{pages}{4945} (\bibinfo{year}{1978}).
  
\bibitem{supp} See http://link.aps.org/supplemental/10.1103/ PhysRevLett.104.227201 for a plot
of the zx and yz contributions to the band structure
around $\Gamma$.  

\bibitem[{\citenamefont{Lee et~al.}(2008)\citenamefont{Lee, Iyo, Eisaki, Kito,
  Fernandez-Diaz, Ito, Kihou, Matsuhata, Braden, and Yamada}}]{iyo08}
\bibinfo{author}{\bibfnamefont{C.~H.} \bibnamefont{Lee}} 
\bibnamefont{et~al.},
  \bibinfo{journal}{J. Phys. Soc. Jpn.} \textbf{\bibinfo{volume}{77}},
  \bibinfo{pages}{083704} (\bibinfo{year}{2008});
\bibinfo{author}{\bibfnamefont{K.}~\bibnamefont{Kuroki}} 
\bibnamefont{et~al.},
  \bibinfo{journal}{Phys. Rev. B} \textbf{\bibinfo{volume}{79}},
  \bibinfo{pages}{224511} (\bibinfo{year}{2009}).

\end{thebibliography}
\end{document}